\documentclass[review]{elsarticle}

\usepackage{hyperref}

\journal{Journal of Magnetism and Magnetic Materials}

\begin{document}

\begin{frontmatter}

\title{Magnetism and charge ordering in high- and low-temperature phases of Nb$_2$O$_2$F$_3$}

\author[imp]{Vladimir~V.~Gapontsev\corref{correspondingauthor}}
\cortext[correspondingauthor]{Corresponding author}
\ead{gapontsevvv@gmail.com}

\author[keln]{ Daniel I.~Khomskii}

\author[imp,urfu]{ Sergey~V.~Streltsov}

\address[imp]{M.N. Miheev Institute of Metal Physics of Ural Branch of Russian Academy of Sciences, 620137 Yekaterinburg, Russia}
\address[keln]{II. Physikalisches Institut, Universit$\ddot a$t zu K$\ddot o$ln, Z$\ddot u$lpicher Strasse 77, D-50937 K$\ddot o$ln, Germany}
\address[urfu]{Ural Federal University, 620002 Yekaterinburg, Russia}

\begin{abstract}
It is shown using {\it ab initio} band structure calculations that Nb$_2$O$_2$F$_3$ is in the orbital-selective regime in the high-temperature phase ($T>90$~K), when two electrons occupy singlet molecular orbital, while the magnetic response comes from the remaining single eleсtron in Nb$^{3.5+}_2$ dimer. The charge order occurs at low temperatures, resulting in the formation of Nb$^{3+}$-Nb$^{3+}$ and Nb$^{4+}$-Nb$^{4+}$ dimers, which makes this system nonmagnetic. The single electron with unpaired spin is transferred to Nb$^{3+}$-Nb$^{3+}$ dimer, but due to a strong splitting of the bonding $xz/yz$ molecular orbitals the low-spin state with $S=0$ is stabilized. We argue that the mechanism of the charge ordering in Nb$_2$O$_2$F$_3$ is the gain in kinetic energy related to the formation of molecular orbitals, which occurs due to a strong nonlinear distance dependence of the hopping parameters.
\end{abstract}

\begin{keyword}
 Dimers, charge ordering.
\end{keyword}

\end{frontmatter}



\section{Introduction}

The $4d-5d$ transition metal (TM) compounds attract considerable attention in recent years. For example, a new Mott state described by the effective total angular momentum J$_{eff}$=$1/2$  and induced by relativistic spin-orbit coupling was found in  $5d$ transition metal oxide Sr$_2$IrO$_4$ \cite{2008prl}. Another interesting effect thoroughly studied nowadays is a possible realization of the so-called Kitaev model \cite{kitaev} on hexagonal lattices, where nearest neighbor exchange Hamiltonian has a form $J\sum_{ij} \vec S^{\alpha}_i \vec S^{\alpha}_j$, where along each of metal-metal bonds only one spin projection $\alpha=x,y,z$ enters \cite{2009prl}. 

Going from the $3d$ to $4d$ and $5d$ transition metal compounds, we not only start to deal with more heavy elements, which implies stronger spin-orbit coupling, but also gradually decrease the on-site Coulomb repulsion (characterized by $U$ parameter) and simultaneously increase spacial extension of $d-$wavefunctions \cite{Khomskii2014}. This drives $4d$ and $5d$ systems to a regime where hopping parameters $t_{ij}$ are comparable with the on-site Hubbard repulsion $U$. Such a growth of $t_{ij}$ is especially pronounced in dimerized systems, where this effect may result in an orbital-selective behavior, when only a part of the $d-$electrons bear magnetic moments, while others form molecular-orbitals \cite{Streltsov2014}. This effect, decreasing the total magnetic moment of a system and suppressing the double exchange mechanism, was found in such systems as Y$_5$Mo$_2$O$_{12}$ \cite{Streltsov2015MISM} and Ba$_5$AlIr$_2$O$_{11}$ \cite{Streltsov2016,Terzic2015}. 

In the present paper we discuss a very similar effect of the orbital-selective behavior and partial suppression of the magnetic moment in a recently synthesized niobium oxyfluoride Nb$_2$O$_2$F$_3$ \cite{Tran2015}. This compound presents a special interest, since here the orbital-selective behavior is observed together with a charge disproportionation. We argue that the mechanism of charge ordering in this system is of a novel type and is connected not with the gain of interaction energy, as in the conventional mechanisms of charge ordering,  but is rather due to the gain in kinetic (bonding) energy due to formation of tightly bound dimers.

Nb$_2$O$_2$F$_3$ crystallizes in a monoclinic structure (space group: I2/a) and undergoes a structural transition at $\sim$90K to a triclinic structure (space group: P-1) in the lower temperatures \cite{Tran2015}. The phase transition at 90K is accompanied by a strong suppression of magnetic susceptibility, which is ascribed to formation of the spin gap. Effective magnetic moment in the Curie-Weiss theory $\mu_{eff}=2.24\mu_B$ (per dimer) in the high temperature (HT) phase is much smaller than expected for Nb having nominal valence $3.5+$.  The transport measurement have been performed for $T>90$K, where Nb$_2$O$_2$F$_3$ was found to be insulating.

\begin{figure}[!b]
\begin{center}
\includegraphics[angle=0,width=0.5\columnwidth]{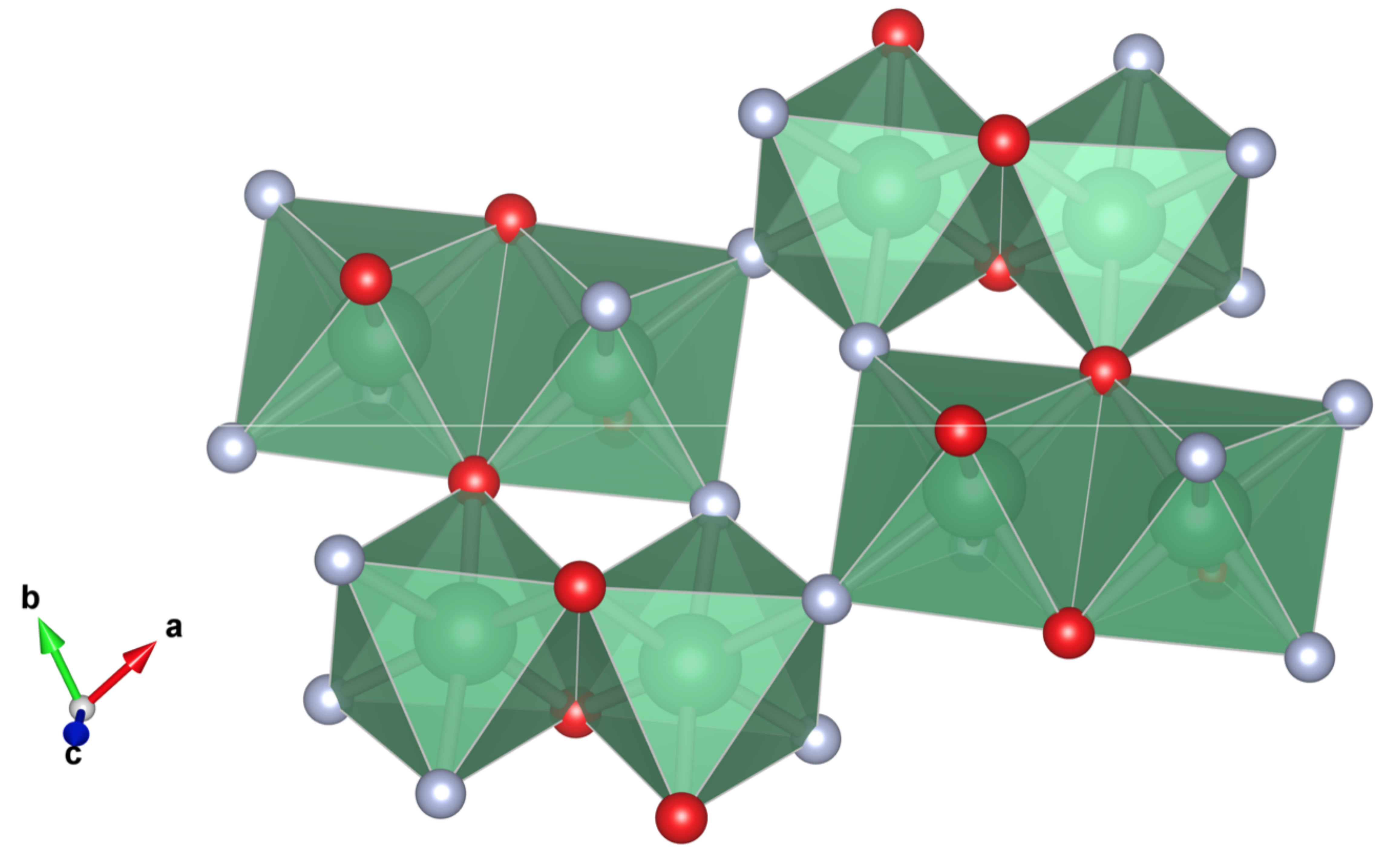}
\end{center}
\caption{\label{cryst.str} Crystal structure of Nb$_2$O$_2$F$_3$ (generated by the Vesta software \cite{vst}). Niobium ions are in the octahedra formed by oxygen (red balls) and fluorine (grey balls).}
\end{figure}

The Nb ions are in octahedral surrounding, and these octahedra share their edges forming Nb-Nb dimers (see Fig.~\ref{cryst.str}). There is one type of such dimers in the HT phase with crystallographically equivalent Nb ions and each Nb having $3.5+$ charge state (electronic configurations $4d^{1.5}$). The transition at $\sim$90K results in a strong modification of the crystal structure. There appears two inequivalent types of Nb and each of them form its own type of dimers: ``short'' Nb1-Nb1 (2.50\AA) and ``long'' Nb2-Nb2 (2.66\AA) dimers. The difference in Nb-Nb bond distances was proposed to be a signature of charge disproportionated state with Nb1 being in the $3+$ (electronic configurations $4d^{2}$) and Nb2 in the $4+$ (configurations $4d^1$) charge states. The origin of the phase transition at 90K is unclear and demands thorough investigation, which we carry out in this paper using ab-initio band structure calculations.

\section{Calculation details}
For the band-structure calculations the pseudopotential PWscf code \cite{pwscf} was mainly used. We utilized the generalized gradient approximation (GGA) with Perdew-Burke-Ernzerhof version of the exchange-correlation potential \cite{pbe} and ultrasoft scalar-relativistic pseudopotentials with nonlinear core correction. The charge density and kinetic energy cutoffs were taken as 40 Ry and 180 Ry, respectively. The mesh of 6x6x6 $k$-points in a full Brillouin zone was used in the calculation.

In order to estimate the crystal field splitting parameters, the Wannier function projection method \cite{Streltsov2005} was used as realized in the linearized muffin-tin orbitals (LMTO) method \cite{lmto}.

All parameters in the paper for both LT (low temperature) and HT (high temperature) phases were calculated with using of experimentally obtained structures \cite{Tran2015}.

\section{Magnetic properties}
We start with the analysis of the nonmagnetic GGA calculation results for the HT monoclinic phase. Since this approximation does not take into account strong on-site Hubbard repulsion \cite{gga+u}, Nb$_2$O$_2$F$_3$ is metallic in this type of calculations with the Nb $t_{2g}$ states forming both valence and conduction bands. The absence of a band gap is a standard drawback of the GGA method, which can be corrected by such methods as LDA/GGA+U \cite{gga+u} or LDA+DMFT \cite{Anisimov97}. In spite of this fact, GGA/LDA is known to be accurate for the 
estimation of the on-site energies and hopping parameters, which are of most interest for us here.

\begin{figure}[t]
\begin{center}
\includegraphics[angle=0,width=0.9\columnwidth]{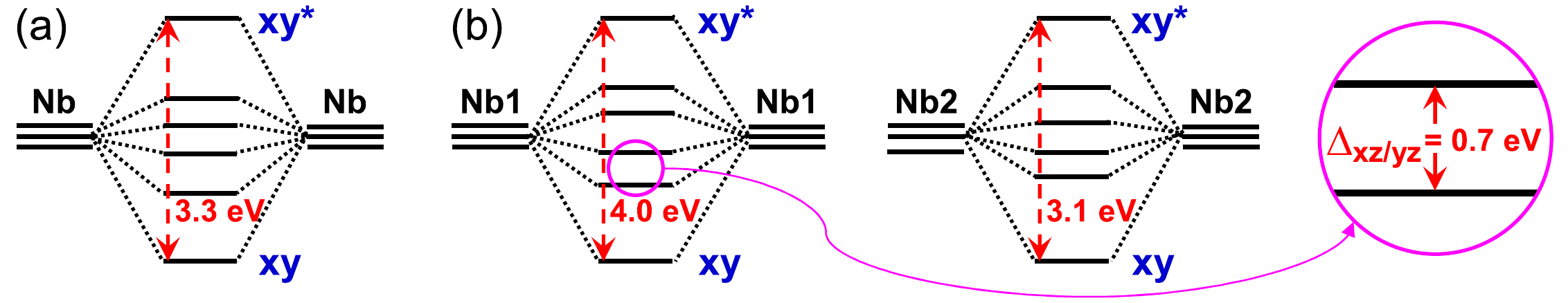}
\end{center}
\caption{\label{hoppings} Formation of the molecular orbitals: (a) in the high temperature and (b) in the low temperature phases, as obtained in of the Wannier function projection method in frames of the LMTO method.}
\end{figure}

We calculated these parameters (hopping integrals and crystal-field splitting) using so-called Wannier function projection procedure. In the edge-sharing geometry there are $xy-$orbitals pointing exactly to each other in the Nb-Nb dimer (we use notations of local coordinate system, where $x$ and $y$ axis are directed to common ligands in a dimer). This results in a strong bonding-antibonding splitting, $\sim 3.3$ eV, see Fig.~\ref{hoppings}. The energy difference between the bonding $xy$ and next (higher in energy) orbitals is $\Delta =$1.1 eV. Using ionic approximation and neglecting on-site Coulomb repulsion one may expect that for a Nb-Nb dimer with 1.5 electrons per site the state with maximal spin $S_{tot}=3/2$ will be stabilized, if $\Delta < J_H/2$, where $J_H$ is the intra-atomic Hund's exchange, but smaller net spin would exist in the opposite case \cite{Streltsov2014,Streltsov2016}. 

We estimated $J_H$ in Nb$_2$O$_2$F$_3$ using the so-called constrained supercell method \cite{Anisimov1991} in the LMTO calculations and found that $J_H = 0.7$ eV. Since the bonding-antibonding splitting is much larger than $J_H/2$, the state with minimal total spin $S_{tot}=1/2$ per dimer is realized in the HT phase of Nb$_2$O$_2$F$_3$, see the level scheme in Fig.~\ref{hoppings}. This state is orbital-selective, since $xy$ orbitals behave here as nonmagnetic molecular-orbital like, while $xz/yz$ orbitals provide local magnetic moment.

\begin{figure}
\begin{center}
\includegraphics[angle=-90,width=0.48\columnwidth]{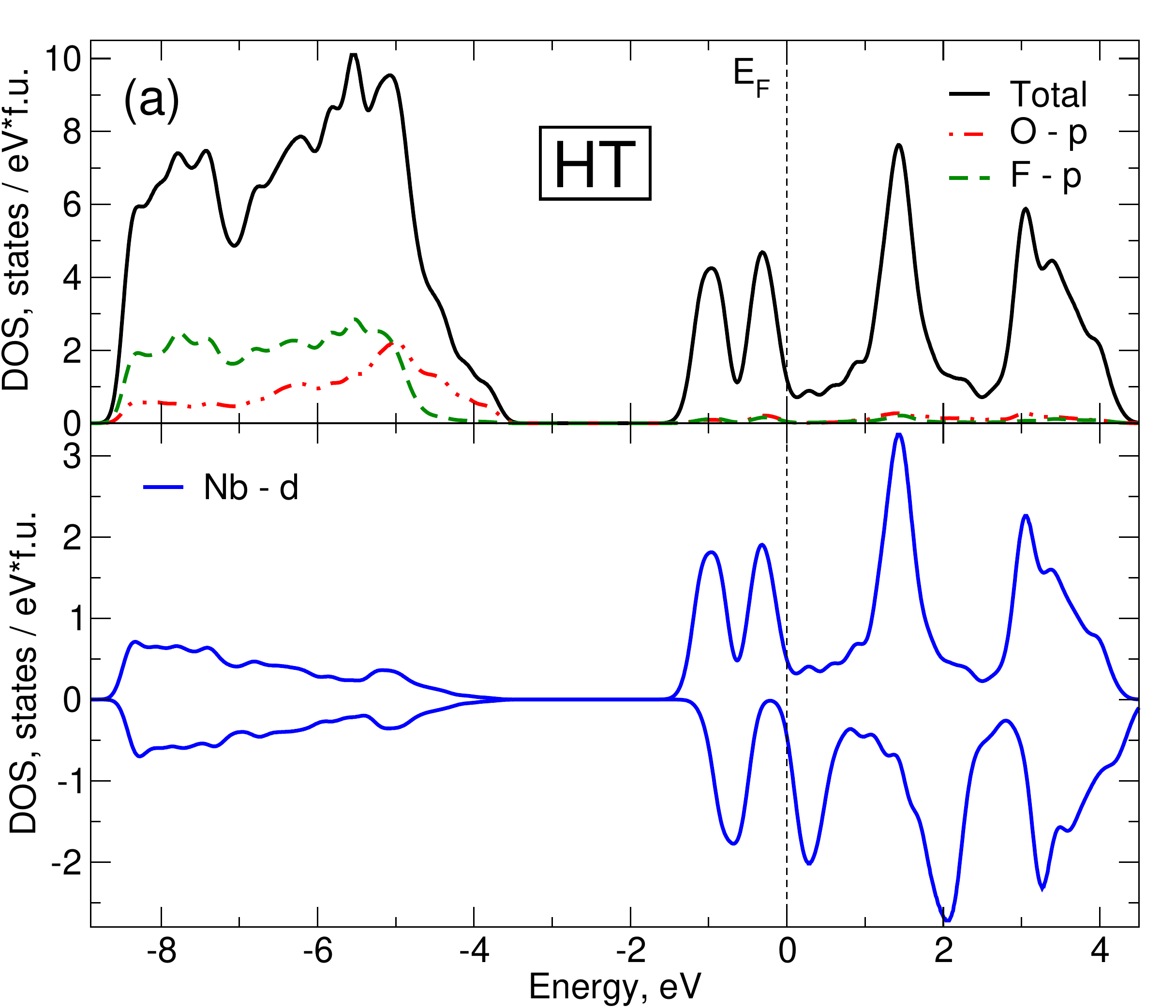}
\includegraphics[angle=-90,width=0.48\columnwidth]{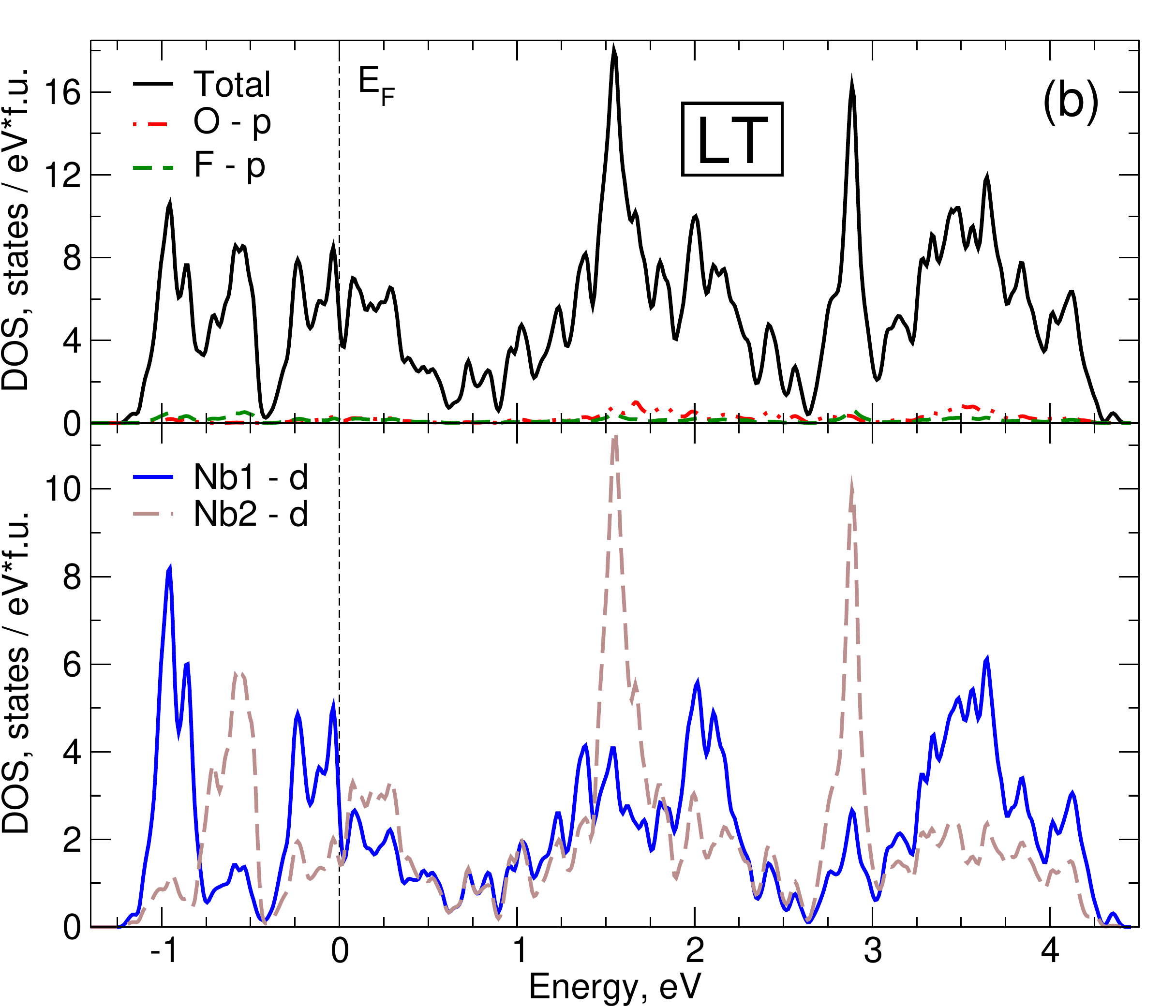}
\end{center}
\caption{\label{DOS} Total and partial densities of states (DOS) for the  high-temperature (HT) monoclinic phase (a) and low-temperature (LT) triclinic phase (b) obtained in the GGA calculation. Only nonmagnetic solution survives in the LT phase, while for the HT phase the results for the ferromagnetic order are presented. Positive and negative values of DOS in lower panel of (a) correspond to different spin projections. The Fermi level is set zero.}
\end{figure}

The ferromagnetic GGA calculations, results of which are presented in Fig.~\ref{DOS}a, show that total magnetic moment is $\mu = 1 \mu_B$ per formula unit (f.u.). This state corresponds to $\mu_{eff} = \sqrt{3} \mu_B$ (per dimer), which is of the order of value observed experimentally \cite{Tran2015}. The bonding and antibonding $xy$ bands are at $\sim -1$ eV and $\sim 3.5$ eV, while remaining peaks in vicinity of the Fermi energy correspond to the $xz$ and $yz$ orbitals. 

The electronic structure in the low temperature (LT) phase is more complex. It was proposed that there can be charge disproportionation in this phase \cite{Tran2015}. Analysis of the hopping parameters shows that the bonding-antibonding splittings in Nb1-Nb1 dimers are even larger than in the HT phase, which is due to smaller bond distance. This is true for both $xy$ and $xz/yz$ orbitals. One may see from the partial density of states (DOS), shown in Fig.~\ref{DOS}b, that both bonding $xy$ and one of the low-lying $xz/yz$ orbitals are occupied in the Nb1-Nb1 dimer. Therefore, one may conclude that there has to be four electrons in the Nb1-Nb1 dimer and only two electrons in the Nb2-Nb2 dimer.
However, difference in L\"owdin charges (total number of $d$ electrons) \cite{Lowdin1950} for two types of Nb is minimal, $\delta n = n_{Nb1} - n_{Nb2} \sim  0.1e$. It is rather typical that the degree of charge disproportionation in a band-structure calculation turns out to be much smaller than what one would expect from formal valencies. The difference in total atomic charges or in corresponding magnetic moments in the GGA or GGA+U calculations is often an order magnitude smaller than it should formally be \cite{Streltsov2014d,Leonov2004,Pickett2014}.

It is clear that, having formally two electrons, Nb2-Nb2 dimer stays nonmagnetic since they occupy bonding $xy$ molecular orbitals. The situation with Nb1-Nb1 dimer with four electrons per dimer is more tricky. In principle one may have three different ``molecular spin-states'' in this case (4 electrons per dimer) with total spin per dimer $S_{tot}=2$, $S_{tot}=1$, and $S_{tot}=0$, and one could even expect magnetic transitions between these states as it occurs in MoCl$_4$ \cite{Korotin2016}. The final electronic configuration is determined by competition between Hund's rule coupling, given by $J_H$, and splitting between molecular orbitals of the $xz$ and $yz$ symmetry. Simple estimates in an ionic model show that these are the states with $S_{tot}=1$ and $S_{tot}=0$, which can be stabilized in the LT phase. If the splitting between bonding $xz/yz$ orbitals $\Delta_{xz/yz} > J_H/2$ (for definition see Fig.\ref{hoppings}) then $S_{tot}=0$ is realized. The Wannier function analysis shows that $\Delta_{xz/yz}=0.7$ eV, which is much larger than $J_H/2 = 0.35$ eV, i.e. for Nb1-Nb1 dimers a nonmagnetic $S_{tot}=0$ state is realized.
This is consistent both with the results of our GGA calculations, where magnetic solution does not survive in the LT phase, and with the experimental results showing nonmagnetic behavior for $T<90$K \cite{Tran2015}. 

\section{Charge ordering}
We saw that the band structure calculations indeed show some degree of the charge disproportionation in the low-temperature phase of Nb$_2$O$_2$F$_3$. Generally speaking there can be many different origins of the charge ordering. The simplest one is the gain in long-range Coulomb (Madelung) energy, like in Wigner crystals \cite{Wigner1983,Hubbard1978}.  In real transition metal compounds the electron-lattice coupling (usually to the  breathing mode of MeL$_6$ octahedra, where L is a ligand ion) often plays even more important role. Sometimes orbital degrees or spin degrees of freedom may help to stabilize the charge order \cite{Khomskii2014,Solovyev1999,Streltsov2014d}.
In all these mechanisms actually the onset of the charge ordered state is related with the tendency to gain an interaction energy.
Using simple arguments we propose that the mechanism of the charge disproportionation in Nb$_2$O$_2$F$_3$ is of very different nature and is connected with the formation of the molecular orbitals in this compound, i.e. it has a two-site nature. 

The bonding-antibonding splitting for the molecular orbitals in Nb$_2$O$_2$F$_3$ is defined by corresponding hopping integrals, or overlap between atomic $d-$orbitals. These hoppings are strongly nonlinear $t \sim 1 /r^5$ \cite{Harrison1999}, where $r$ is the distance in a dimer. Therefore making short and long dimers instead of two Nb-Nb dimers having the same bond length we win a lot of kinetic energy. Similar idea was used before to explain isotope effect in manganites \cite{Babushkina1998} and cobaltates\cite{Babushkina2014eng}. 
Moreover, it turns out that one gains a kinetic energy not only due to the $xy-$orbitals having direct overlap in the common edge geometry. Other Nb $4d-$orbitals also contribute to lowering of the charge ordered state. One can construct molecular orbitals of the $xz/yz$ symmetry. By making a linear combination $xz+yz$ we maximize direct hopping keeping overlap via ligand $p_z$ orbital nearly the same \cite{Kimber2014}. 

One may estimate the total energies of the charge ordered and homogeneous states in a simple ionic approximation using Kanamori parametrization for the Coulomb interaction \cite{Kanamori1963} and calculated with the LMTO hopping parameters. Since the Nb-Nb distance in the HT phase is nearly exactly the arithmetic mean of short and long Nb-Nb bonds in the LT phase, one may use these data to calculate the difference in hopping parameters for homogeneous (H) and charge ordered (CO) states $\delta t_{xy}$ and $\delta t_{xz+yz}$ for the $xy$ and $xz+yz$ orbitals, correspondingly. Then the total energy difference between these two states in the ionic approximation can be estimated as
\begin{eqnarray}
\delta E = E_{H} - E_{CO} = \delta t_{xy} + \delta t_{xz+yz} - U/2.
\end{eqnarray}

The direct calculation based on the results of the Wannier function analysis shows that  $\delta t_{xy}=0.8$ eV and $\delta t_{xz+yz}=1$ eV. One may expect that $U$ for Nb is smaller than 3 eV (typical value for, e.g., Ru$^{4+}$ having much larger number of electrons \cite{Lee2006}). Thus, we see that at zero temperature indeed a gain in the kinetic energy due to formation of molecular orbitals exceeds a loss in the Coulomb energy because of the charge ordering, which in its turn is even overestimated in the simple Hubbard-like treatment used above: when we take into account  $p-d$ covalency, the actual charge transfer in transition metal system with charge ordering is never complete, as was assumed in the crude estimate above,  but is   much smaller, typically near $ \pm0.2e$ \cite{Khomskii2014}. At finite temperature $T$ instead of the total energies, $E$, one has to compare free energies $F=E-TS$ and increasing temperature the entropy term will dominate, which stabilizes homogeneous state and finally leads to the transition to this state.  

However, one has to note that there are two important ingredients, which we did not take into account in aforementioned model. First of all, Coulomb interaction not simply adds a penalty due to inhomogeneous charge distribution, but also modifies the wavefunctions of the corresponding states increasing contributions from the ionic terms (i.e. changing them from molecular-orbital like to Heitler-London like). This will decrease energy gain due to the formation of the molecular orbitals \cite{Streltsov2016}. 

\begin{figure}[t]
\begin{center}
\includegraphics[angle=0,width=0.6\columnwidth]{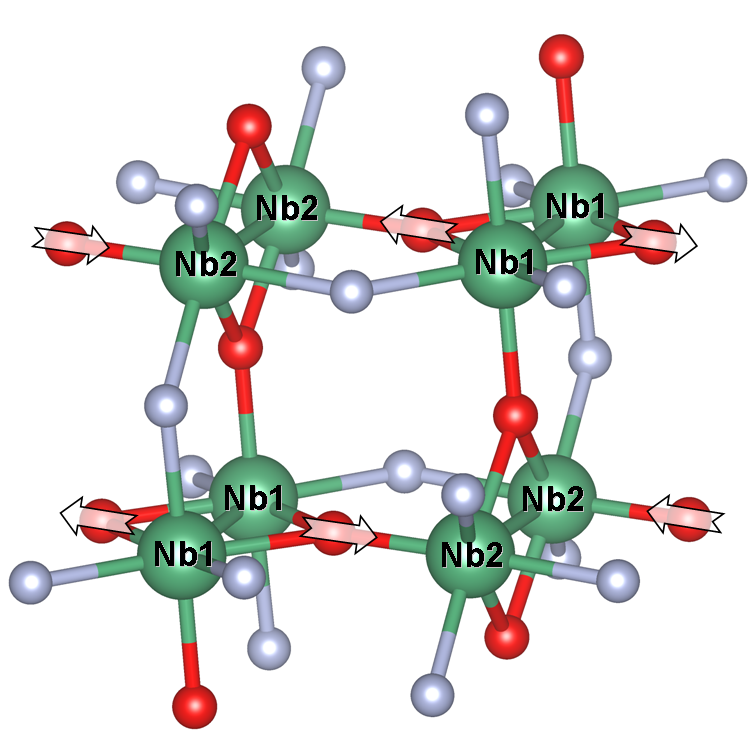}
\end{center}
\caption{\label{Distortions}  Illustration of cooperative distortions, which support formation of the charge ordered state.}
\end{figure}

Second, there is a very important contribution due to the elastic energy, which normally stabilizes homogeneously states. This correction is not easy to estimate on a model level for such a complicated structure as in Nb$_2$O$_2$F$_3$. However, there is an extra feature of the Nb$_2$O$_2$F$_3$ crystal structure, which has to be mentioned and which helps the observed charge ordering. Dimerization of one class of Nb drives another one to change its charge state. Indeed, if, e.g., we decrease Nb1-Nb1 bond distance and want to keep Nb1 valence as it is (i.e. keep the same average Nb1-O and Nb1-F bond lengths), then oxygen ions linking Nb1-Nb1 and Nb2-Nb2 dimers should move towards Nb2 as it is shown in Fig.~\ref{Distortions}. This should lead to decrease of corresponding Nb2-O bond distance and hence to the increase of the valence of Nb2 as a result. It is interesting to note that the low temperature crystal structure of Nb$_2$O$_2$F$_3$ indeed shows this effect with shortest Nb-O bond (1.89~\AA, while other Nb2-O bond lengths are 2.03~\AA~and Nb2-F are $\sim$2.07-3.0~\AA) linking two dimers together. Thus, we see that in this sense dimerization goes hand in hand with the charge disproportionation in Nb$_2$O$_2$F$_3$. This situation actually strongly resembles that in M$_1$ and M$_2$ phases of VO$_2$ \cite{Ghedira1977,Whittaker2011,Khomskii2014}. 

\section{Summary}

To sum up, we show in the present paper that magnetic response in Nb$_2$O$_2$F$_3$ at  the high-temperatures ($T>$90 K) is related to the orbital selective regime, when part of the electrons form molecular orbitals while other electrons have local magnetic moments. The charge disproportionation, which occurs at $T\sim$90 K is seen in the GGA calculations, but its degree ($\delta n \sim 0.1$ electron) is far from what one would expect from naive expectations based on the formal ionic valences. The mechanism of the charge ordering is argued to be related with a sizable kinetic energy gain due to formation of two molecular orbitals in short Nb$^{3+}$-Nb$^{3+}$ dimers caused by a strong nonlinearity of the distance dependence on electron hopping. We think that this mechanism of charge ordering, stabilized not by decrease of  interaction energy, but rather by the gain  in kinetic energy, may be operative in many other systems, especially consisting of structural dimers.

\section*{Acknowlegments}
We are grateful to B. Lorenz for informing us of their results on Nb$_2$O$_2$F$_3$. This work was supported by Russian Science Foundation via project 14-22-00004.

\section*{References}

\bibliography{bibfile}

\end {document}